# A Cost-Aware Mechanism for Optimized Resource Provisioning in Cloud Computing

Safiye Ghasemi[1], Mohammad Reza Meybodi[2], Mehdi Dehghan Takht Fooladi[2], and Amir Masoud Rahmani[1,3]

[1]Computer engineering department, Science and Research Branch, Islamic Azad University, Tehran, Iran.
[2]Computer Engineering and Information Technology, Amirkabir University of Technology, Tehran, Iran.
[3]Computer Science, University of Human Development, Sulaimanyah, Iraq.

*Abstract*— **Due to the recent wide use of computational resources in cloud computing, new resource provisioning challenges have been emerged. Resource provisioning techniques must keep total costs to a minimum while meeting the requirements of the requests. According to widely usage of cloud services, it seems more challenging to develop effective schemes for provisioning services cost-effectively; we have proposed a novel learning based resource provisioning approach that achieves cost-reduction guarantees of demands. The contributions of our optimized resource provisioning (ORP) approach are as follows. Firstly, it is designed to provide a cost-effective method to efficiently handle the provisioning of requested applications; while most of the existing models allow only workflows in general which cares about the dependencies of the tasks, ORP performs based on services of which applications comprised and cares about their efficient provisioning totally. Secondly, it is a learning automata-based approach which selects the most proper resources for hosting each service of the demanded application; our approach considers both cost and service requirements together for deploying applications. Thirdly, a comprehensive evaluation is performed for three typical workloads: data-intensive, process-intensive and normal applications. The experimental results show that our method adapts most of the requirements efficiently, and furthermore the resulting performance meets our design goals.**

*Index Terms*—**Cloud computing, cost, learning automata, resource provisioning, services, virtual machine.**

## 1. Introduction

Cloud computing is a technology that provides various services such as Infrastructure-as-a-Service (IaaS) and Software-as-a- Service (SaaS) via related providers [1, 2]; these services are provided to cloud users as a pay-per-use basis model. Nowadays, enterprises prefer to lease their required services such as applications from related providers as online services instead of buying them as on-premise ones [3], especially in the cases that the services are not needed for long use. Thus, a large number of applications that were running on users' desktops are transformed to SaaS services which run on the IaaS cloud providers [4]. The automated provisioning of these applications has many challenges [5, 6]; it is a significant step to satisfy the quality of services (QoS) which can increase the satisfaction of users as well. The cost of provided services is a dominant part of the providers' and users' satisfaction factor [7] which makes resource provisioning strategies more critical. The server cost is the most effective metric in the price of provided services [8]; as the hosting infrastructure costs are considered as the largest share of provisioning cost [9], cloud systems put the main burden on providers, and hence how to reduce the costs is an urgent issue for providers while provisioning. None of current cloud provisioning mechanisms provides cost-effective pay-per-use model for SaaS applications [1].

IaaS is a computational service model that has extra computing resources, such as processing power, storage, and network bandwidth [10] to provide resources to SaaS providers by the aim of virtualization technologies, which is one of the core technologies of cloud computing. Virtual machines (VMs), which are used to host the requested applications, share the available physical resources provided by IaaS providers [11]. The problem of VM placement has introduced as a crucial problem [10, 2, 12]. Optimizing the process of resource management is an effective way of decreasing service costs [8, 13, 14]. Resource management related problems include resource allocation, resource adaptation, resource brokering, resource discovery, resource mapping, resource modeling, resource provisioning and resource scheduling; these are discussed in [14]. Some of approaches that tackle this crucial issue include following. Dynamic SLAs mapping are considered to restrict the number of resource types [15] in a way to optimize costs with respect to the public SLA template. Resource provisioning approaches in cloud systems are addressed in different researches. In [4], provisioning of continuous write applications in cloud of clouds is proposed. In [16], cost-aware solutions for dealing with MapReduce workloads are discussed. To make an optimal provisioning decision in [10], the price uncertainty from cloud providers and demand uncertainty from users are taken into account to adjust the tradeoff between on-demand and over-subscribed costs. Optimizing the resource provisioning task by shortening the completion time for the users' tasks while minimizing the associated cost is performed by dynamic resources provisioning and monitoring approach in [17]. Moreover, the approach includes a new virtual machine selection algorithm called the host fault detection algorithm. Finally, in [9] a cooperative resource provisioning solution for four heterogeneous workloads: parallel batch jobs, web servers, search engines and MapReduce jobs are introduced. Live migration [18] as an important component of cloud computing paradigm provides extreme versatility of management. It is applied to load balancing approaches [19] to improve resilience and availability of services [20]; but it comes at a price of degraded service performance during migration. Transferring the memory image of service from the source host to the destination host is considered as migration process. It could be evaluated by computing the expected degradation in service level due to the bandwidth limitations while migrating services between VMs for load balancing. The cost of live migration process consists of determining when to migrate, deciding what services to migrate, determining where to migrate these services, the pre-copy phase cost, the down time cost, and cost of the amount of bandwidth to be used for the migration in each step [19, 20]. We do not address this issue as our main goal is to efficiently manage the cost of provide the cost request.

It is to be noted that the existing optimization provisioning approaches for cloud solutions cannot deal with the application demands; besides, they are cost-aware resource provisioning approaches based on per-job optimization without considering different attributes of individual requirements for VMs placement. In this paper, we apply a learning-based provisioning approach that can allocate available VMs of SaaS providers of cloud to application demands. The providers, which denote SaaS providers in this article, deploy the applications on their particular platforms for providing to demanding users [21]. As providers are unaware of the conditions of requests, learning automata (LAs) are used in the process of provisioning to tackle the lack of information. We propose variable structure LAs in provisioning process of each provider to handle requests. According to the fact that applications comprised of different services, like security services, database services and etc, LAs find the optimal combination of VMs for hosting each service of the demanded application. All services of an application must be deployed on proper VMs to enable execution of the application. Thus, it is necessary to have the minimum requirements of each service before an application provisioned. In other words, if we have a request of $n$ applications $App_i$, $i$=1..$n$, each with $s_i$ services then the resource provisioning approach must consider the total requirements of $\sum_{i=1}^{n} s_i$ services. Considering such property for applications makes us to care about this structure in comparison with two-tier structure of workflows. A resource provisioning approach for a workflow which comprises of $n$ processes and dependencies between them produces the order of $n$ processes based on their dependencies [22]. The required resources of processes are allocated based on the order determined by the scheduler; it is not needed to consider the requirements of all processes together [17], while in provisioning of an application there is no order between its services and they must be deployed based on their minimum requirements. Therefore, the provisioning approach of an application searches among available VMs and finds the most proper VM for each service of the application, which is performed by LA.

The main aim of this article is to represent a dynamic mechanism that facilitates the optimized resource provisioning process by the use of LAs. The unique benefits of our optimized resource provisioning (ORP) approach are as follows. First, unlike existing models that allow only workflows in general form of jobs and their dependencies, we provide a cost-effective resource provisioning solution for applications by considering the fact that, each application comprises of different services; the main challenge is to provision totally required services of an application. Secondly, ORP is a learning automata-based approach, which selects the most proper computing resources in form of VMs for hosting each service of the demanded application. It considers both cost and computing requirements, as the formalization of measures, for deploying applications based on their attributes; these are applied to LAs to assess the performance evaluation of the approach. Finally, comprehensive evaluations are performed for three typical application types: data-intensive, process-intensive and normal applications. The simulations of ORP present its adaption to most requirements efficiently, while reducing the expected costs, and furthermore the resulting performance meets our design goals as well.

The rest of this paper is organized as follows. Section 2 presents the body of the article, i.e. the system model and assumptions of cloud computing environment. Section 3 formalizes the resource allocation problem and introduces the proposed algorithm. The experimental setup and simulations for the performance evaluation of ORP are described in Section 4. Finally, Section 5 summarizes and concludes this article.

## 2. System model

The proposed optimized resource provisioning approach, named ORP, significantly improves cost-effective issues of providing the cloud services to users in form of applications. Cloud providers deploy the demanded applications of users on their particular infrastructures [21]. It is to be noted that a provider does not know the upcoming requests in cloud environment. Therefore, it must make decisions based on current situations without any accurate long-term decisions. Thus, a decision maker is required to overcome such limits of variable and unknown situations. Our proposed approach makes optimal provisioning decisions based on current conditions by the use of LAs. As the process proceeds, the provider performs the optimal provisioning decisions based on the requests. According to current requests, the provider determines its way of resource provisioning to maximize its profit while satisfying users. In this section we firstly, describe primitives of learning automaton in Section 2.1; then, in Section 2.2, the proposed provisioning mechanism is generally presented; finally, the details of the mechanism is discussed, i.e. the performance factor of ORP while using LAs is formulated in 2.3.

### *2.1 Learning Automata*

Learning automaton [23] is an automatic learning model which its learning relates to the way of collecting and using knowledge during its decision making. The learning process of each learning automaton has three main components: the *LA*, the *Environment* and the *Reward*/*Penalty* structure. They are briefly explained as follows.

1) LA: The LA can be modeled in form of a quintuple as $\{Q, \alpha, \beta, F(.,.), H(.,.)\}$ [23], where:

- $Q$ is a finite set of internal states of LA as $\{q_1, q_2, ..., q_z\}$, where $q_t$ is the state of LA at instant $t$.
- $\alpha$ is a finite set of actions of LA as $\{\alpha_1, \alpha_2 ... \alpha_r\}$, where $\alpha_t$ is the action that the automaton has performed at instant $t$; note that $\alpha$ is the output of LA.
- $\beta$ is a finite set of replies of the environment after that the LA applies the action; $\beta = \{\beta_1, \beta_2, ..., \beta_m\}$ where $\beta_t$ is the response of the environment at instant $t$; in other words, it is the input of LA.
- $F$ is a mapping function that maps the current state and the input of LA to the next state, i.e. $Q \times \beta \rightarrow Q$.
- $H$ is a mapping function that generates an action according to the current state and the input of the LA, i.e. $Q \times \beta \rightarrow \alpha$.

2) Environment: An environment is the medium in which the automaton functions. The environment can be mathematically modeled in form of a triple as $\{\alpha, \beta, C\}$ [23]; $\alpha$ and $\beta$ are the set of inputs and outputs of the environment, respectively; $C$ is a set of penalty probabilities that the environment considers for each of its inputs as $\{c_1, c_2, ..., c_r\}$.

LA interacts with the environment in a feedback loop, as depicted in Fig. 1; in this interaction, the input of LA is the output of the environment and vise versa. The environment replies to the LA based on the selected action. LA updates the probabilities of its actions according to the environment responses. Updating is performed with a particular reinforcement scheme; the negative values of reinforcement indicate punishment, and positive values express reward.

There are several models of LA defined based on the response set of the environment. Models in which the output of the environment can take only one of two values, 0 or 1, are referred to as P-models. In such case, the value of 1 corresponds to an unfavorable response which means failure or penalty, while output of 0 denotes the action of LA is favorable. A further generalization of the environment, called Q-models, allows finite response sets that take finite number of values in an interval [a, b]. When the output of the environment is a continuous random varia-

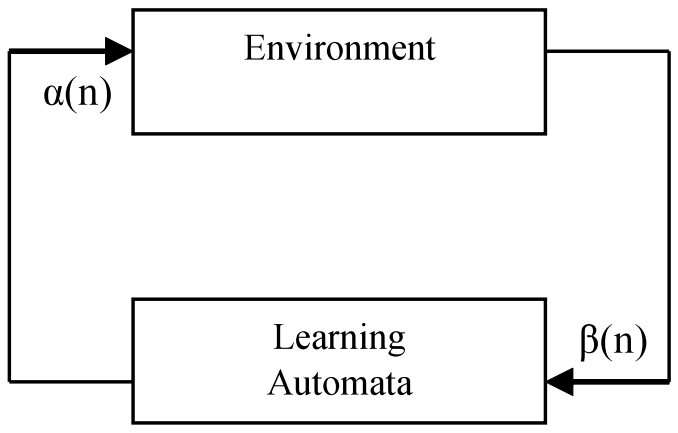

**Fig. 1** Interaction of a Learning automaton with an environment [23]

-ble with any possible values in an interval [a, b], is named S-model.

Each LA uses the following equations to update the probabilities of its action set after receiving replies of the environment; for desired replies Eq. (1) is used and for undesirable replies, Eq. (2).

$$\begin{aligned} p_i(n+1) &= p_i(n) + a.(1 - p_i(n)) \\ p_j(n+1) &= p_j(n) - a.p_j(n) \forall j \; j \neq i \end{aligned} , \quad (1)$$

$$\begin{aligned} p_i(n+1) &= (1-b).p_i(n) \\ p_j(n+1) &= \frac{b}{r-1} + (1-b).p_j(n) \forall j \; j \neq i \end{aligned} . \quad (2)$$

Where, $p_i$ denotes the probability of selected action $i$; the parameter $a$ is associated with reward response, and the parameter $b$ is associated with penalty response; $r$ is the number of available actions of LA. The equations are written in a way to let the probabilities in interval [0, 1] and $\sum_{i=1}^{r} p_i = 1$ is true.

LAs are used in problems faced by an agent that learns behavior through trial-and-error interactions with a dynamic environment [23]. It has proved effective behaviours in situations which the state of the environment is unknown and changes very quickly. In systems-science literature, learning automata are among the models that are employed successfully to tackle the problem of decision making under uncertainty [24]. The probabilities of taking different actions would be adjusted according to their previous successes and failures [25].

### *2.2 Provisioning Mechanism*

The model of our considered cloud is derived from [8]. In this model, cloud comprises of users who demands for applications, SaaS providers who provide the demanded application of the users on their leased VMs, and IaaS providers who provide virtual resources in form of VMs to SaaS providers. The operation of a cloud lies with the cloud provider. The cloud model enables the users to have a computing environment without investing a huge amount of money to build computing infrastructures. According to the considered cloud market model in [8, 15, 26], our proposed resource provisioning scenario comprised of users, SaaS providers (providers), and IaaS providers as well; the model is depicted in Fig. 2. Users demand applications from a provider. The provider hosts a set of applications on its virtual infrastructures, named virtual machines (VMs). Before the requests provided, computing resources have to be provisioned from providers. IaaS providers package the resource requirements of providers into VMs [10, 2, 3], and then services can be deployed on VMs in the same way as physical machines [3]. This way of provisioning enables independent services [27]. In other words, IaaS providers offer requested VMs to the SaaS providers. SaaS providers can buy new VMs while resource provisioning. This assumption causes some delays while provisioning; the delay is because of the time that IaaS providers require for preparing new VMs to SaaS providers, which is discussed in [16], with details.

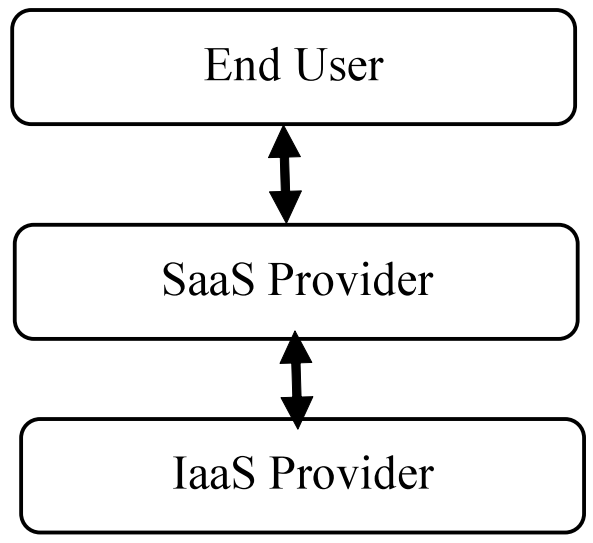

**Fig. 2** Cloud computing market model [4]

A user may demand different applications from a provider; in fact, users simply submit their requests of applications [10, 16]. Furthermore, they can specify some quality concerns which is known as service level agreement (SLA). There are a large number of commercial providers which may enter cloud market and offer a number of different types of applications [15]. It is clear that users choose providers which provide the application with the less price and acceptable performance. On the other hand, as providers pursue the profit, they try to attract as more users as possible, and thus, they must offer affordable prices with satisfactory performance. The providers have complete control on the attributes of virtual resources that are devoted to the requests. In this paper we have focused on resource provisioning process of providers to minimize the infrastructure cost by minimizing the cost of VMs which are required to handle the requests. Obviously, the providers must decrease the costs to have acceptable prices for services without losing the profits.

The proposed resource provisioning model, which is applied to providers of the considered cloud market, is presented in Fig. 3. The depicted model of Fig. 3 is a SaaS provider, such as what presented in Fig. 2, by omitting users and IaaS providers; instead of users and IaaS providers of Fig. 2, a *Request Pool* and a *Virtual Resources* frame is considered, respectively. The Request Pool gets the requests from users; the Virtual Resources frame gets VMs from IaaS provider and stores in the provider for hosting the applications.

As mentioned before, the provider's goal is to maximize its profit while providing satisfactory services to the users with affordable prices and acceptable performances. The proposed approach of this research reaches the goal by descending the infrastructural costs which is performed by Provisioning System (Fig. 3). According to [8, 9], the server cost contributes the largest proportion of the total cost of provisioning services. Users determine their demanded software requirements, e.g., operating systems and applications [3]; they specify the performance of their requests by some predefined parameters which are determined by SLA. The provisioning system gets the requests, which are stored in Requests Pool, by communicating with Requests Interface. Request Interface is placed under Requests Pool in the model depicted in Fig. 3. Previously mentioned, applications are hosted on VMs, which are provided on a pay-per-use basis by IaaS providers. Each application comprises of several services; the application run by the aim of these services. As an instance, a CRM application, which is provided by different vendors such as Salesforce.com or Microsoft Dynamics, may need some services such as database, security, calculating and accounting to be able to run. Requests are related to an application demands, and they are formalized in form of applications as follows

$$Req = \langle AppID, s, Srv, \tau \rangle. \quad (3)$$

Where, *Req* denotes a request stored in Request Pool; *AppID* is the identification of the application which is requested in *Req*; *s* is the number of services of which the request comprised; the list of the services of the demanded application of *Req* is stored in *Srv*, which is modeled as $Srv=[VMSrv_1, VMSrv_2, \ldots, VMSrv_s]$; finally, $\tau$ denotes deadline of the application which is determined by the user. The services can be located on different VMs based on the *VMSrv* determined by *Srv*. Each of these services is supposed to be hosted on an individual VM. Since VMs have different properties, cloud providers have a limited set of available configurations [12]. A set of such configuration, which is determined by *VMSrv* in *Srv*, includes following properties as <*VM type*, *Core*, *Memory*, *Storage*, *Throughput*, *Hour cost*>, e.g., a user may request a VM as <*'large', 3, 30 MB, 2048 MB, 100KB/s, 3.400$*>. We consider the following formulation for introducing a VM, which is stored in Virtual Resources layer of providers,

$$VM_{ij} = \langle Size_{ij}, Core_{ij}, Memory_{ij}, Storage_{ij}, Throughput_{ij}, Hourcost_{ij} \rangle. \quad (4)$$

$VM_{ij}$ presents the properties of VM *i* stored in Virtual Resources of provider *j*. $Size_{ij}$ depicts type of $VM_{ij}$, which can be small, medium, large, …. $Core_{ij}$, used to presents the number of cores that $VM_{ij}$ has; it is the computing power of $VM_{ij}$. $Memory_{ij}$ and $storage_{ij}$ are mostly in Giga Bytes (GB). $Throughput_{ij}$ denotes read/write network throughput of $VM_{ij}$; The cost of $VM_{ij}$ per hour is $hourcost_{ij}$. The presented configuration of a VM introduced in Eq. (4) is based on Amazon EC2. List of VMs stored in Virtual Resources layer of provider *i* is called $VMList_i = \{VM_{i1}, VM_{i2}, \ldots, VM_{ir}\}$; *r* is the number of VMs available in the Virtual Resources of the provider.

The requests, stored in Requests Pool, are forwarded to Provisioning System via Request Interface; the Request Interface is supposed as a gateway that handles the requests and the replies. The provisioning system decides to accept a request if there are sufficient resources available in Virtual Resources of the provider based on the parameters of the requests; if the provider accepts a request it must specify the most appropriate VMs to deploy it.

### *2.3 Performance Factor*

The proposed provisioning system depicted in Fig. 3 has to specify the list of VMs that would better host the application. The process operates by the means of a learning scheme consisting of LAs. We used variable structure learning automata (LA) in each SaaS provider to find the best virtual machines (VMs) for deploying the current request including its services. The LA updates the probabilities of taking the actions via learning algorithms to reach the best values of probabilities for current request. As the requests and also the available VMs in the provider change very quickly using LA to find the best VMs for each request is a useful approach in this scenario that can lead each automaton quickly converges to its optimal action. LA explore between the limited numbers of states including placement of services of the request on VMs of the provider.

Analyzer plays the role of the environment described in Section 2.1; Analyzer can communicate with Resource Manager and Request Interface to get informed of the available Virtual Resources of the provider and properties of the current request. The actions of LAs, $\alpha$, is the available VM of Virtual Resources. LA comprises LAs which select VMs according to the probabilities of the action set for each of the services in the requested application of the request. The selected VMs, $\alpha_t$, is passed to Analyzer as the action of LA at instant $t$. Then, Analyzer evaluates the action of LA to generate a reinforcement signal as its output. This reinforcement is firstly computed the performance factor named $\rho$. Then, it is evaluated by comparing with a predefined threshold. Since the proposed LAs are P-model one, thus, if $\rho$ is less than the thresholds, the selected VM is rejected by the learning system; otherwise the learning system accepts the selected action.

Suppose that LAs in LA select VMs 1 to $s$ for hosting services 1 to $s$ of the demanded application in the current request. LA passes $\alpha_t = VM_1 \ldots VM_s$, at instant $t$, to Analyzer; Analyzer evaluates the action by comparing the selected VM, with the requirements of service $j$; the result of the comparison is considered as the performance factor $\rho$, depicted in Eq. (5).

ρ=▯,▯ ▯▯▯▯▯

▯▯▯=k=15▯▯▯−▯▯▯▯▯▯▯6, f▯▯ ▯▯▯ &▯
∄▯▯▯−▯▯▯▯▯<0 0 , &▯▯▯▯▯▯▯▯▯ (5)

$\rho^i_j$ is computed by considering both the offered computing resources of the allocated VM and the cost of the selected VM. It divides the summation of adaption of chosen VM with the related service by the cost of the chosen VM. $p_{ik}$ denotes property $k$ of VM $i$ with the related property $k$ of service $j$ in the demanded application which is named as $Ser_{jk}$ in Eq. (5). The considered properties of VMs and services are size, memory, storage, core, and Throughput of VMs as $k = 1 .. 5$, respectively. $p_{i6}$ denotes per hour cost of VM $i$; in other words, $p_{i6}$ introduces the 6$^{th}$ property of VM $i$ as depicted in Eq. (4).

$\rho^i_j$ computes the performance factor for each service of the demanded application separately, as each service may be deployed on an individual VM. Analyzer which is aware of both the state of request via Request Interface and the available resources of the provider via Virtual Resources plays the role of the environment of LA in Provisioning System. LA updates the probabilities of its action set based on reinforcement signals of Analyzer. The update is computed based on the result of comparison of $\rho$ of the selected VM with the threshold. Then the environment passes the rejection or acceptation of the particular VM to LA. The interaction of LA and the environment continues until the learning is converged (the probabilities of actions remain unchanged for a while or reach to predefined values), or the number of iterations reaches a maximum limit. It is obvious that the selected VM is omitted from action set after being chosen for the current service. Then, the process is repeated for next service of the application by reinitializing the values. Finally, a list of VMs is generated which are chosen for deploying the services of the requested application. The list is sent to Request Interface.

Based on Eq. (5), VMs which have more compatibility with the requirements of the services and have less price must have more chance to be chosen by LA in subsequent runs of the provisioning system. The algorithm of resource provisioning system depicted in Fig. 3 is studied in details in following section.

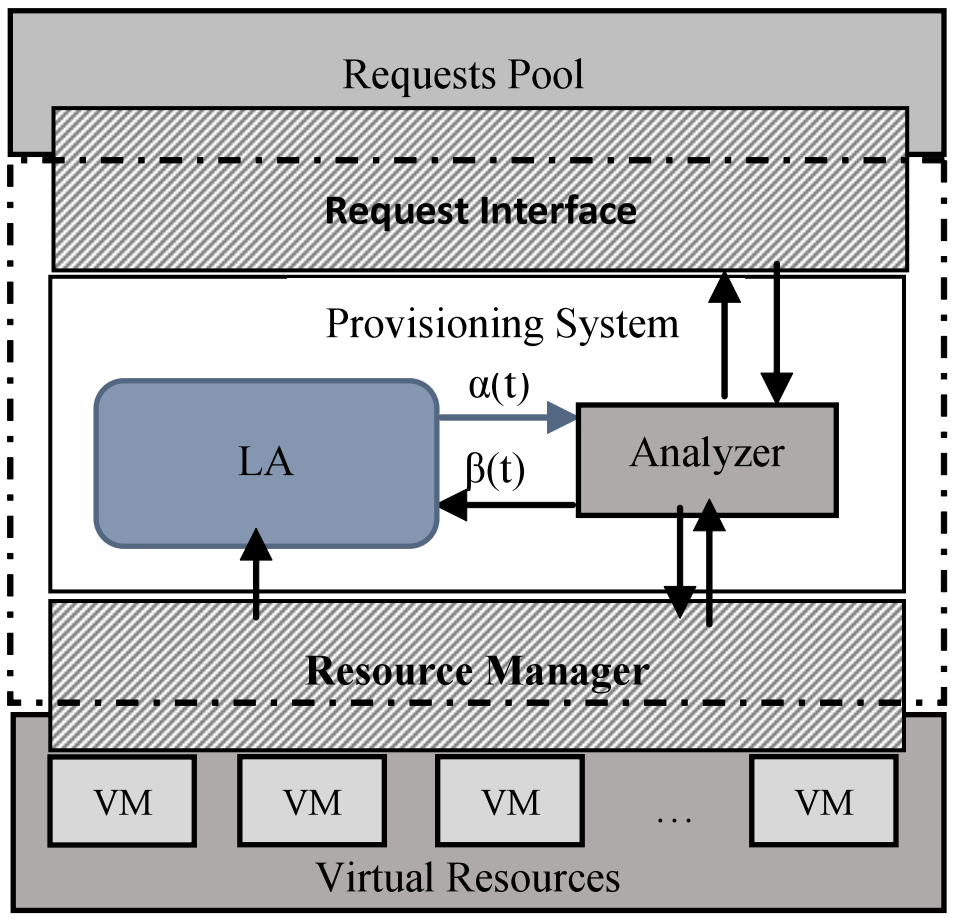

**Fig. 3** The model of resource provisioning optimization considered in a SaaS provider

## 3. The Provisioning Mechanism Formulation

As previously mentioned, our proposed provisioning system uses LAs to find the most proper VM for hosting each service in the request. As mentioned before, after receiving a request by Request Interface, it is forwarded to Provisioning System; each request, *Req*, consists of *s* services and Provisioning System must run LAs for the services in *Req*. It is to be noted that LAs search in *VMList* of the provider and find the best VM for each service. Thus, the problem involves finding *s* VMs between *r* VMs of different properties existed in Virtual Resources of the provider, $1 \leq i \leq r$, where each VM is available in a certain size $VMSrv_i$, as introduced in Eq. (4) . The problem is to fill the requirements of *Req*, which consists of *s* fixed services, with *SelectedVMList* = [$VM_1$, ..., $VM_s$] to yield a minimal value for the costs while mapping the requirements. Firstly, Provisioning System must solve the problem by deciding on which service would be better to host at first. Then, it must find the best VM for hosting that service. Since the aim of ORP is to optimize the performance and the cost of provisioned resources in cloud market, performance and cost compatibility model is defined in form of Eq. (5) to get the goal. The equation takes the attributes of the selected VM and the attributes of the request of the user as input variables to help to quantify the performance and the cost. The details of this process are presented in this section.

Previously mentioned, a provider must offer satisfactory levels of performance guarantee for deploying demanded applications. In addition to the performance concerns, profit of the provider is a considerable factor in resource provisioning and providing the requested applications as well. Therefore, requests of users are the main revenue source of providers. The mechanism of our proposed provisioning system is applied in order to obtain the optimal VMs selection to increase the providers' profits. The reason is that, ORP finds the nearest VMs to the requirements of services with the lowest prices, which causes an optimal utilization of resources.

Initially, the probabilities of actions of LAs are the same, signifying that any of the VMs is equally the same to be selected by the provisioning system. Thus, VMs are randomly selected; then, Analyzer calculates the performance of the selected action and sends a reinforcement signal to LA. The probability of the selected VMs is increased when the environment sends a favorable response to LA, i.e. if VM *i* of the provider is chosen for service *j*, then $\rho_j^i$ is better than the threshold value; the probabilities of other VMs are decreased as well. Otherwise, the probability of the selected VM is reduced, while the probabilities of other VMs are increased; the response of the environment is unfavorable. Eqs. (1) and (2) are used for updating the probabilities of actions of LA.

The provisioning strategy is presented in Algorithm 1. Table 1 summarizes key notations used in the algorithm. The providers, which can deploy the requested application on their VMs, run the algorithm. A provider can deploy the request if it has sufficient virtual resources based on the requirements of the demanded application; otherwise, the provider can buy new VMs from IaaS providers to provide the request. In this case, Virtual Resources of the provider is changed; thus, $\alpha$ is changed as well. In other hand, the user has to suffer a delay for virtual resources preparation which might not be acceptable in comparison with the time required for other providers to provide his/her request. As mentioned before, a large number of commercial providers currently exist in cloud market, offering a number of different types of applications [15].

The algorithm of Provisioning System consists of the process of interaction of LA and Analyzer (Figs. 1 and 3). The provider runs Algorithm 1 after receiving a request. At the end of Algorithm 1, a list of VMs, named *SelectedVMList*, is sent to Request Interface (Fig. 3) as the output of the algorithm. Firstly, the provider checks whether its available virtual resources, called VMs, can tackle the current request, *Req*, or not; in case that it cannot tackle (i.e. ~*Tacke*(*VMs* , *Req*) in Line 5) which means that the provider requires additional resources, negotiating with IaaS providers via function *Negotiation*(*IaaS*) starts. Finally, after provisioning new virtual resources, they are added to the current VMs in Line 7, by function *Renew*(). Then the main part of provisioning of requests starts; the algorithm runs for each service, named *Srv*, of the demanded application in *Req*, from Line 10. Besides, the process is executed in a loop which iterates until the model is converged (Lines 9-29) for each service in *Req*. The model is converged if the probability of a selected VM exceeds 95%, or $\rho$ remains unchanged for several iterations. Otherwise, the process stops when it iterates for a maximum limit. When the probability of a VM converges to 1, then the selected action of LA is optimal and therefore the selected VM is the one that must be a part of the output of Provisioning System. In Line 12, function *select(VMList)* randomly selects a VM from the list of virtual resources of the provider named *VMList* based on the probability of VMs; this selection is the action of LA. Then, in Lines 14-19, the selected action is evaluated to help to generate a reinforcement signal in Lines 24-27. Function *Adapted*(*p*,*q*) compares the values of *p* and *q*, then, returns their difference as the compatibility of *p* to *q*; *p* relates to the attributes of the selected VM and *q* relates to the attributes of the considered service. The considered compatibility parameters *size*, *core*, *mmry*, *strg*, and *trgp*, which are size, memory, core, storage, and throughput, respectively, denote the compatibility of attributes of the allocated VM to the attributes of the considered service of the request. Then, in Line 19, a variable, named *Total*, is computed according to the values of compatibility parameters. The coefficients $v_1$, $v_2$, …, and $v_5$ balance the compatibility parameters *size*, *core*, *mmry*, *strg*, and *trgp* based on the type of the requested application; e.g. in a data-intensive application the storage, memory and throughput are more important than the other factors. In other words, these coefficients enable us to compute the summation of compatibility parameters, which have different types. *Total* computes the compatibility of the chosen VM to the considered service based on all attributes of the requirements.

Table 1
Parameter definitions and their values

| Variable | Description |
|---|---|
| *Req* | A request in Request Pool of the provider |
| $Srv_i$ | Service *i* in the list *Srv* of *Req* |
| $VMSrv_i$ | Infrastructural requirements of service *i* |
| *s* | Number of services that *Req* comprises of |
| $VM_i$ | VM *i* in Virtual Resources of the provider |
| $VM_i.Size$ | Type of $VM_i$ |
| $VM_i.Memory$ | Memory of $VM_i$ |
| $VM_i.Core$ | Number of cores that $VM_i$ has |
| $VM_i.Storage$ | Storage of $VM_i$ |
| $VM_i.Throughput$ | read/write network throughput of $VM_i$ |
| $VM_i.HourCost$ | Cost of $VM_i$ per hour |
| $\rho_j^i$ | Performance factor of selecting $VM_i$ for deploying $Srv_j$ |
| *b* | Parameter of penalty |
| *a* | Parameter of reward |

After assessing the chosen VM, in Line 20, the algorithm computes the performance factor presented in Eq. (5). The *Normalize*(*p*) function in Line 20, converts the performance factor into the range of 0 and 1, as follows,

$$Normalize(p)=\frac{p-A}{B-A}. \quad (6)$$

*p* is the main value of performance factor before being normalized; *A* and *B* are the minimum and the maximum values that *p* may take, respectively. The values of *A* and *B*

**Algorithm 1** Algorithm of Resource Provisioning system

The algorithm run by the providers which can deploy the services of requested application on their VMs.

```
1   Inputs:
2       Req
3   Output:
4       SelectedVMList
5   if ~Tackle(VMs , Req)
6       Negotiation(IaaS);
7       VMs = Renew();
8   end
9   repeat
10  foreach srv in Req_i.Srv do
11      repeat
12          SelectedVMList[srv] = Select(VMList);
13          // starts to evaluate the action of the chosenVM
14          size = Adapted(SelectedVMList[srv].Size , srv.Size);
15          mmry = Adapted(SelectedVMList[srv].Memory , srv.Memory);
16          core = Adapted(SelectedVMList[srv].Core , srv.Core);
17          strg = Adapted(SelectedVMList[srv].Storage , srv.Storage);
18          trgp = Adapted(SelectedVMList[srv].Throughput , srv.Throughput);
19          Total = v1×size + v2×mmry + v3×core + v4×strg + v5×trgp;
20          ρ[srv, SelectedVMList[srv]] = Normalize
                    (Total/SelectedVMList[srv].HourCost);
21      until convergence occurs
22  end foreach
23  ρ = Sum(ρ); // according to Eq. (5)
24  if ρ < b then
25      Update the probabilities according to Eq. (1)
26  if ρ > a then
27      Update the probabilities according to Eq. (2)
28  until convergence occurs
29  end
```

are calculated based on the minimum and the maximum values of VMs in *VMList* of the provider, respectively.
After normalizing the performance factor of the selected VM, updating the probabilities of the available actions is performed in Line 25 and 27 for favorable selection and unfavorable selection, respectively. As mentioned before, favorable and unfavorable selections are determined by comparing with some thresholds *a>0* and *b<1* which are determined according to the model in the experiments.

## 4. Performance Evaluations

In this section ORP is evaluated in terms of its economical resource provisioning decisions; the performance is compared with variety of systems. Section 4.1 introduces the setting of simulated cloud market environment in a quantitative manner with descriptions of parameters setting. In Section 4.2, firstly, the efficiency of learning system used by provisioning system is analyzed; then, some comparisons with other approaches are discussed.

### 4.1 Experimental Setup

In this section, firstly the local test bed of experiments including the architecture of simulation model, the status of the providers, VMs, IaaS providers, and requests are introduced; then the parameters of LAs are discussed.

#### 4.1.1 Local test bed environment

We have modeled providers of cloud computing in a discrete event simulation, for evaluating performance of the proposed approach. The simulation model is shown in Fig. 4; it consists of a unit for arrivals of requests which is named Request Generator and forwards the requests to the Request Pool of providers, IaaS provider which provides infrastructural requirements of providers in form of VMs, and several SaaS providers equipped with different resource provisioning techniques which have two outputs, one for evaluating the performance of LAs and one for Comparison

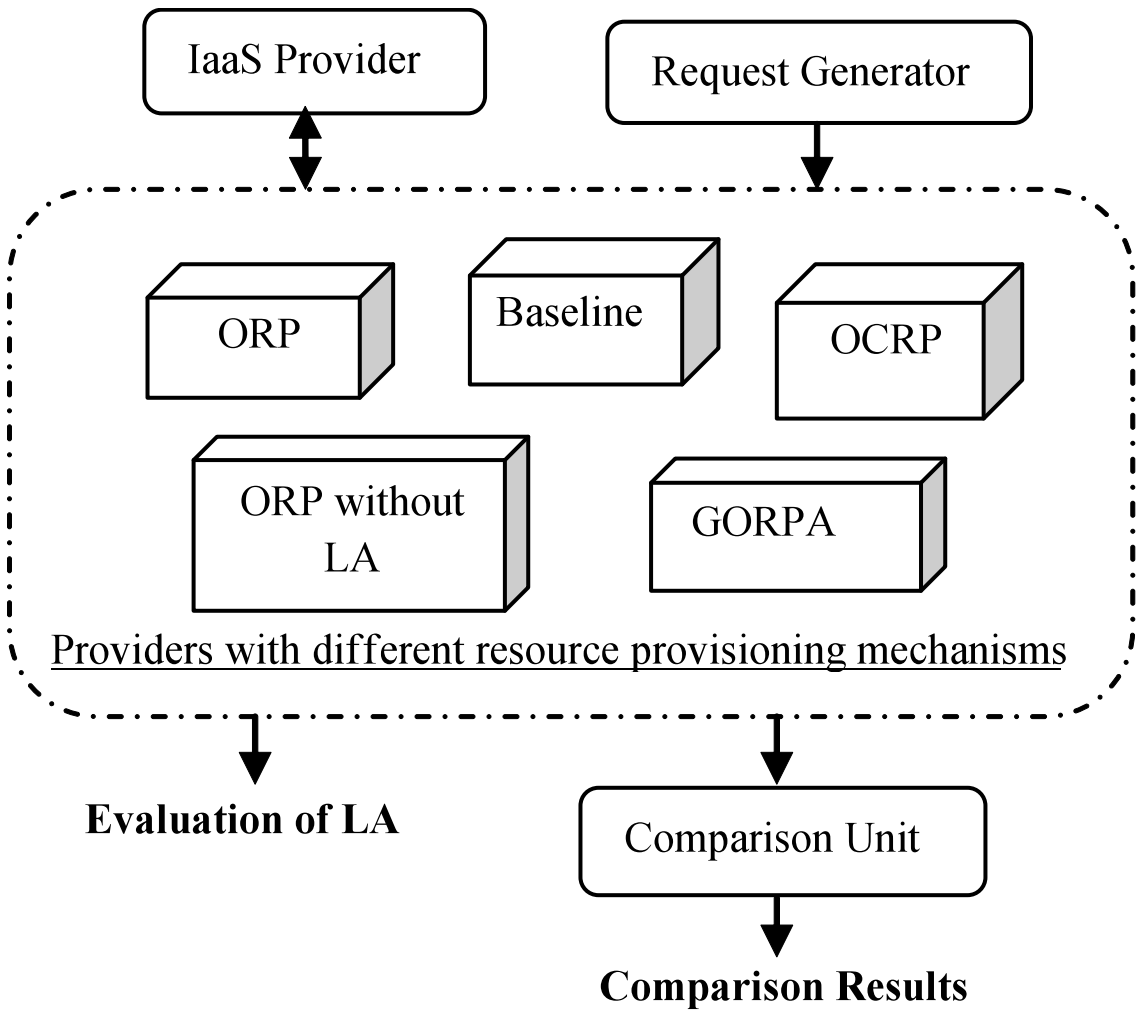

**Fig. 4** Simulation model

Unit which compares ORP approach with other provisioning techniques applied to other SaaS providers. These techniques include GORPA [4], OCRP [10], and ORP without LA which selects VMs randomly.

For the sake of simplicity, we have assumed that IaaS providers offer the computing resources to the available providers with configurations of instances of Amazon EC2 (Feb 2017) depicted in Table 2. There are a rapidly growing number of SaaS providers which provide required applications of users [11]. Users can easily find the latest list of SaaS providers offering software solutions in their interested area. For our experiments, 15 providers are defined as the ones which use ORP approach for resource provisioning. In particular, when the algorithm converges, the required statistics are calculated to indicate the behavior of ORP on average. These providers initially own predefined number of VMs with configurations of VMs depicted in Table 2; the number of VMs exist in each provider is a random variable determined by uniform distribution (20, 50). It is assumed that IaaS provider offers unlimited amount of resources in terms of VMs, so the simulations are not face with shortage of VMs. As mentioned in Section 3, let *VMList* denotes the set of VMs available in the provider. It is assumed that each VM hosts a distinct service of the request of application (e.g., some VMs for database services and another for an individual web service); however, the solution can be also extended by omitting this assumption. It is to be noted that certain amount of physical resources is required for hosting provided VMs of IaaS providers. The VM instance is determined according to the required amount of resources of a VM [10]; some instances are presented in Table 2. The prices, in Table 2, are defined in dollars per resource unit for an hour in Windows usage.

In our experiment, no probability distribution of arrival of requests is considered since they are stored in Request Pool and they are sent to providers per request. We use Grid Workloads Archive (GWA), GWA-T-12 Bitbrains, from Delf University (http://gwa.ewi.tudelft.nl) as our workload traces [28]. It contains the performance metrics of 1,750 VMs from a distributed datacenter from Bitbrains.

Table 2
Properties of VMs with the prices

| Attr. / Size | VCPU | Memory (GB) | Storage (GB) | Price per VM/$ |
|---|---|---|---|---|
| t2.small | 1 | 2 | 1×4 | \$0.026 /Hour |
| t2.medium | 2 | 4 | 1×4 | \$0.052 /Hour |
| m3.medium | 1 | 3.75 | 1×4 | \$0.070 /Hour |
| m4.large | 2 | 8 | 1×32 | \$0.1041 /Hour |
| c3.large | 2 | 3.75 | 2×16 | \$0.141 /Hour |
| c4.xlarge | 4 | 7.5 | 2×40 | \$0.2067 /Hour |
| c4.2xlarge | 8 | 15 | 2×80 | \$0.412 /Hour |
| r3.large | 2 | 15 | 1×32 | \$0.175 /Hour |
| i3.large | 2 | 15.25 | 1×32 | \$0.109 /Hour |
| i3.xlarge | 4 | 30.5 | 1×80 | \$0.218 /Hour |
| i3.2xlarge | 8 | 61 | 1×160 | \$0.436 /Hour |

Bitbrains is a service provider that specializes in managed hosting and business computation for enterprises. Each file of GWA-T-12 contains the performance metrics of a VM. In our experiments fastStorage is applied.

The workload is entered to the model from Request Generator. GWA consists of different VM requirements of 1750 requests. Our evaluations use some of these requests and they are stored in Request Generator; they are sent to the providers in an offline manner one per request, and they are stored in Request Pool of the provider. The format of each request of application in GWA is compatible with the introduced metrics in Eqs. (3) and (4).

As previously mentioned, the requests are in form of application; applications are software packages which consist of different services, e.g., operating system, database, and any other utility service. For simplicity we consider 20 types of applications, in GWA, that a user can demand. The licenses of applications are assumed to be purchased from software vendors by the providers; thus, users, instead of buying licenses, desire lease an application from the providers to save their budgets. The leased applications are needed to be hosted on the proper virtual resources of providers [10], named VMs. There is not large number of different types of VMs offered by providers; for instance, Amazon introduces only few derivations of their basic resource type [15]; Table 2 depicts offered instances of VMs in our experiments. The proper VMs are the ones which are compatible with the requirements of the demanded application. The user pays for the application according to the license cost of the application per running VMs [10]. The cost of a licensed application is determined based on the selected VMs. The required VMs of each service of the application are varied following GWA.

#### *4.1.2 Sensitivity analysis*

First, we start with a sensitivity analysis on the learning parameters $\alpha$ and $\beta$, in order to study their effects on the performance of the ORP and also to find the best value of them. To reach this, an input set with 50 requests is considered. Two main parameters of ORP algorithm are the reward and the penalty parameters of LAs, $\alpha$ and $\beta$, respectively. To achieve more certainty and to accelerate the convergence of the algorithm, these parameters can be varied from 0.7 to 0.9 and from 0 to 0.1, respectively in different experiments and the reported values are depicted in Fig. 5. The vectors of the chart in Fig. 5, which are marked with $\alpha$, $\beta$ and *Iteration* labels, indicate the reward and the penalty parameters of LAs and the average number of iterations required for convergence of LAs in 10 different providers, respectively. Fig. 5 shows that the best value for $\alpha$ and $\beta$, which has the least number of iterations, is 0.8 and 0.05, respectively.

Furthermore, the number of steps of LAs must be determined in order to stop the algorithm when the other convergence conditions are not accessed. According to Fig. 5, it can be concluded that LAs find the solution and converge after approximately 200 steps, in average.

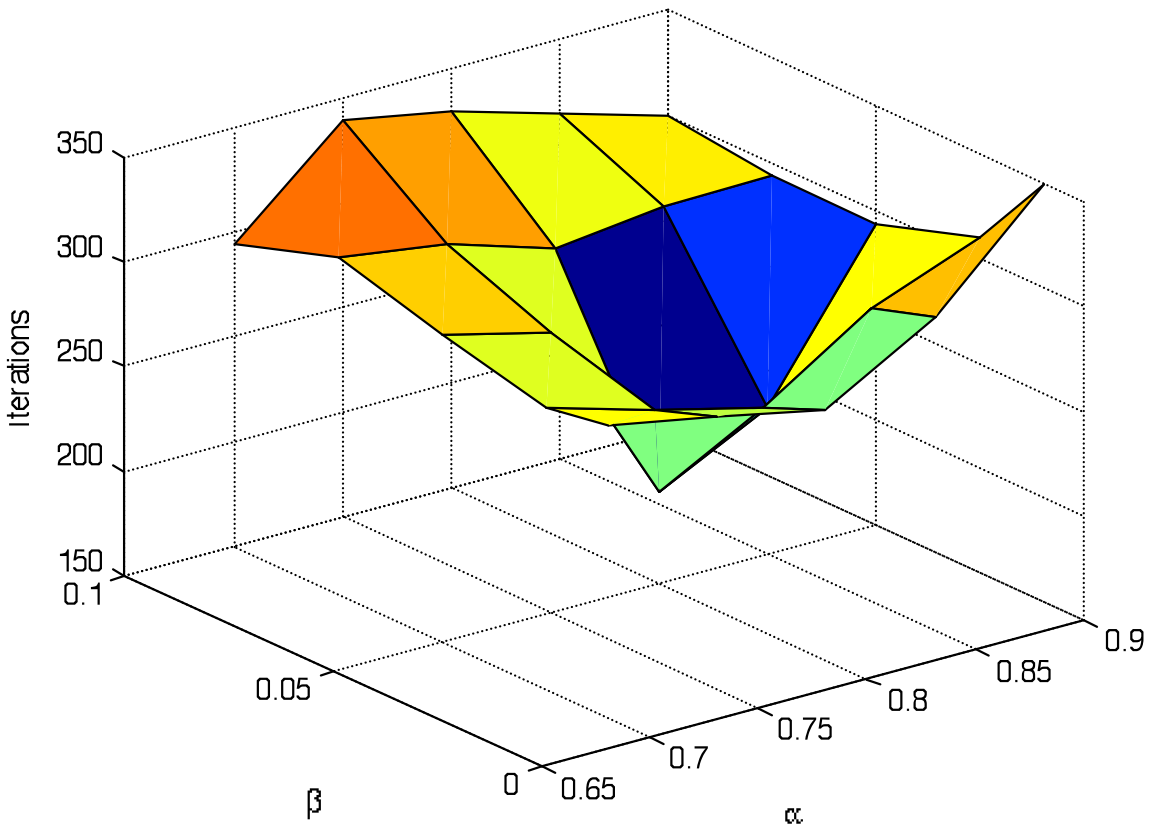

**Fig. 5** The probability distribution of VMs during iterations of LA

### *4.2 Experimental results and analysis*

At the beginning of the simulation, the generated requests are sent to Request Pool. Then the requests are forwarded to Request Interface of the providers. Each provider tries to find the best combination of VMs for hosting the requests. At the end of the processing of all requests, some statistical data such as average number of rejected requests, utilization of VMs and cost of the provisioned resources are generated. We compare the performance of resource provisioning mechanisms depicted in Fig. 4, including ORP, GORPA [4], OCRP [10], and ORP without LA which selects VMs randomly. The results of the experiments are included a baseline as well. The input of baseline experiments is a set of predefined requests which is sent to Request Pool. Our performance evaluation is measured according to three performance metrics: the number of requests that are denied to be processed by the approaches as no VM is available; in this case the algorithms run without adding new VMs. The second evaluated metric is utilization of VMs, and the third one is the total provisioning cost. The experiments are designed to evaluate the values of these metrics in form of the outputs of Comparison Unit, in Fig. 4; Section 4.2.1 discusses the results of the evaluations of these metrics. In addition, a comprehensive evaluation of ORP is performed for three typical workloads: data-intensive, process-intensive and normal applications; it is discussed in Section 4.2.2.

#### *4.2.1 Comparison with other resource provisioning mechanisms*

The experiments use the values of parameters shown in Tables 1 and 2, with the same workload traces from GWA, for all provisioning approaches depicted in Fig. 4.

It is to be noted that for evaluating the throughput and the QoS violation of our provisioning approach, a fix number of VMs is supposed to enable the comparison of the approaches more accurately [4]. However, in other experiments that performed for validating the evaluations of the proposed approach, such as costs and utilization, SaaS providers can take advantage of using VMs in an elastic fashion.

We firstly compare the throughput and the QoS violation of ORP in comparison with GORPA introduced in [4], OCRP in [10], and LA-omitted ORP approaches of provisioning. In these two experiments, the same amount of virtual infrastructures in form of VMs is assumed for providers based on Table 2. Although the whole point of using VMs is to virtualize the infrastructure, and to request and release VMs on demand in an elastic fashion to adapt to the workload but in these experiments, it is assumed that the providers cannot take advantage of this in order to verify the throughput of the approaches and the average number of requests which cannot be processed by the approaches. Requests are classified into three classes depending on their resource requirements as in [4]: small, medium and high demand classes; some of the requests applied to the experiments are presented in Table 3, derived from GWA-T-12 Bitbrains.

The results of this comparison are depicted on the bar chart of Fig. 7. It is shown that the throughput of ORP is more than other approaches. The reason is that in ORP, LAs find the most proper VMs amongst Virtual Resources of the provider; thus, the compatibility of the attributes of the VMs and the requests are well considered and many of the requests can be provisioned with a determined amount of virtual resources. While LA is used for finding the most proper VMs for a request there are little number of requests which cannot be processed and are rejected. The performance is better than both of GORPA and OCRP; GORPA is designed for continuous write applications and considers the shortest path in terms of data transmission cost between VMs. OCRP provisions based on reservations which does not performs well in these experiments in comparison with ORP.

Table 3
Properties of requests used for experiments of Fig. 7

| Request | Services | VCPU | Memory (GB) | Storage (GB) |
|---|---|---|---|---|
| Req Class$_1$ | VMSrv$_1$ | 1 | 1 | 1×4 |
| | VMSrv$_2$ | 1 | 4 | 1×4 |
| Req Class$_2$ | VMSrv$_1$ | 2 | 4 | 1×4 |
| | VMSrv$_2$ | 2 | 8 | 1×32 |
| | VMSrv$_3$ | 4 | 8 | 2×40 |
| Req Class$_3$ | VMSrv$_1$ | 2 | 15 | 2×32 |
| | VMSrv$_2$ | 4 | 15 | 2×80 |
| | VMSrv$_3$ | 4 | 30 | 1×32 |
| | VMSrv$_4$ | 8 | 15 | 1×32 |
| | VMSrv$_5$ | 8 | 30 | 1×80 |

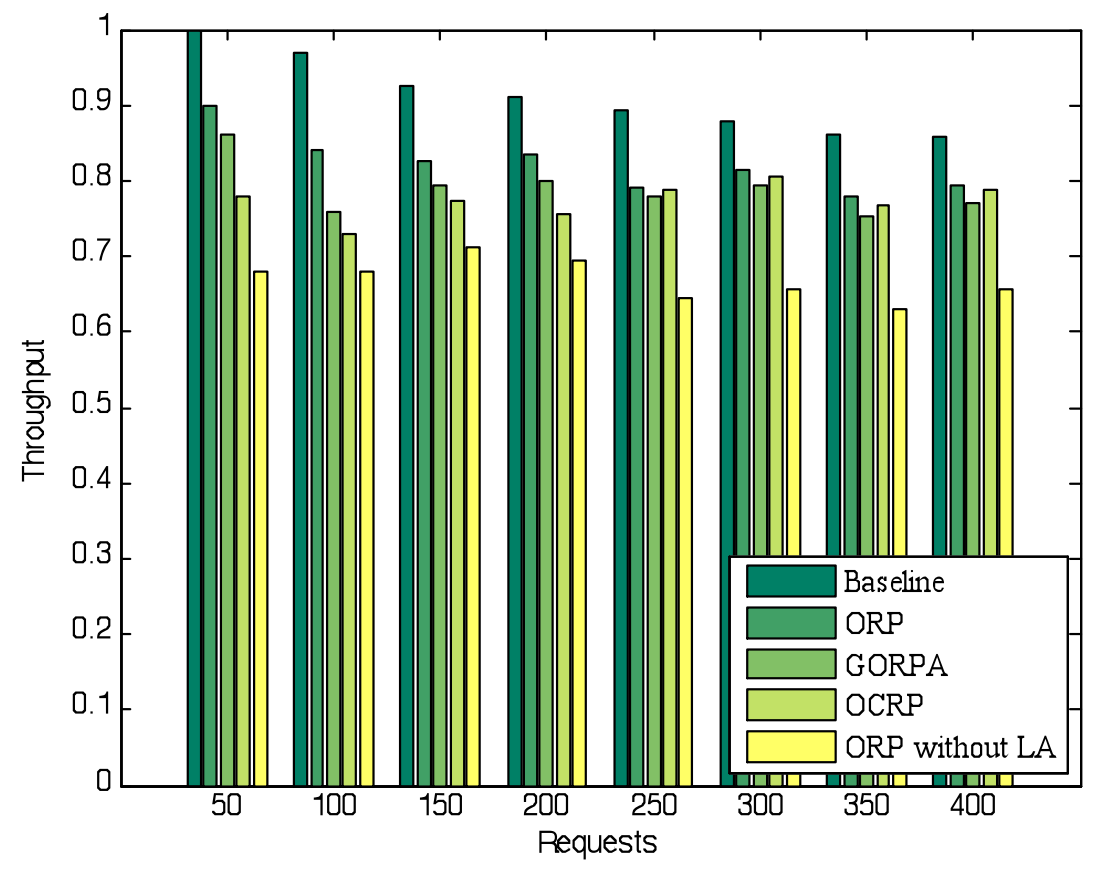

(a)

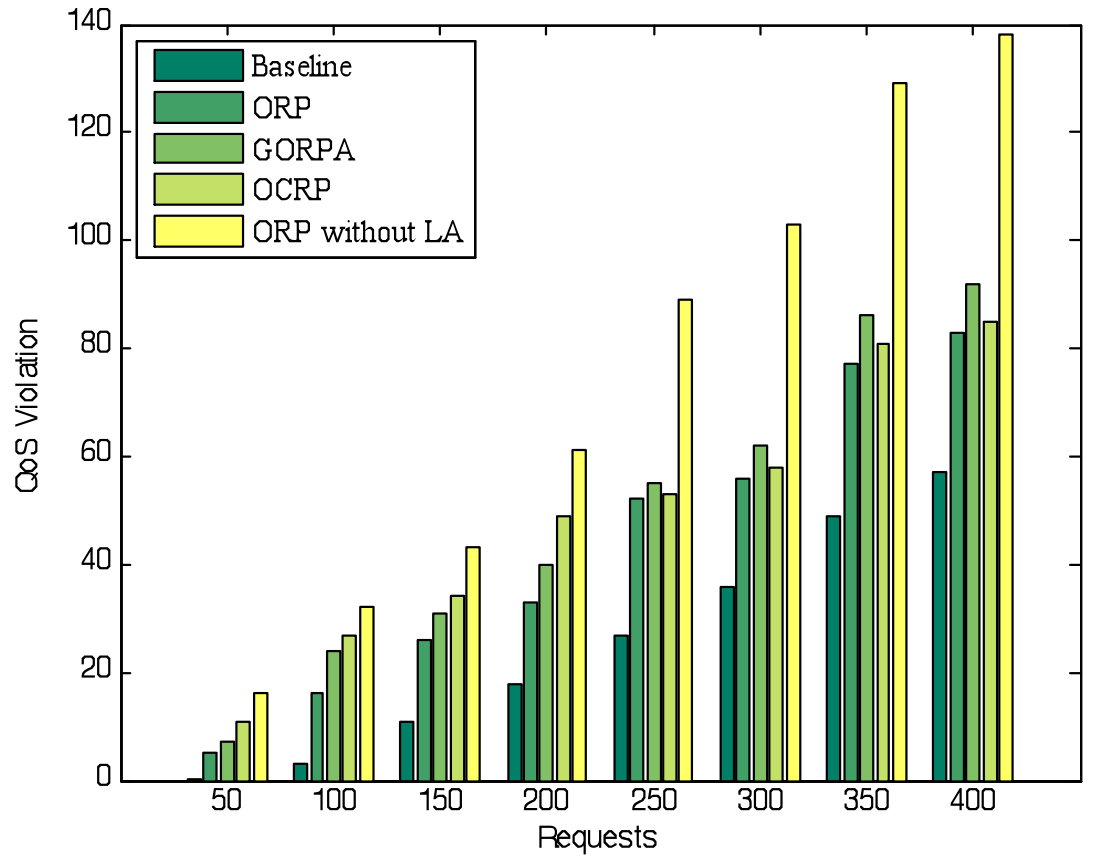

(b)

**Fig. 7** QoS evaluation (a) Throughput (b) number of rejected requests

Utilization is another parameter which is used to evaluate the performance of the proposed approach. To compute the utilization of VMs of a provider the average CPU, memory, and disk for the set of VMs for each request is used.

It is to be noted that if a provider allocates VMs to requests without considering the requirements then the utilization of the provider will be a low value; on the other hand if the compatibility of the requirements of the requests and the selected VMs is high then the provisioning approach will have a great utilization of its virtual resources as well.

In this section we compare the utilization of virtual resources of providers while using ORP approach and OCRP [10], GORPA [4], and ORP without LA. The results of this comparison, which are generated by Comparison Unit of simulation model depicted in Fig. 4, are represented in Fig. 8.

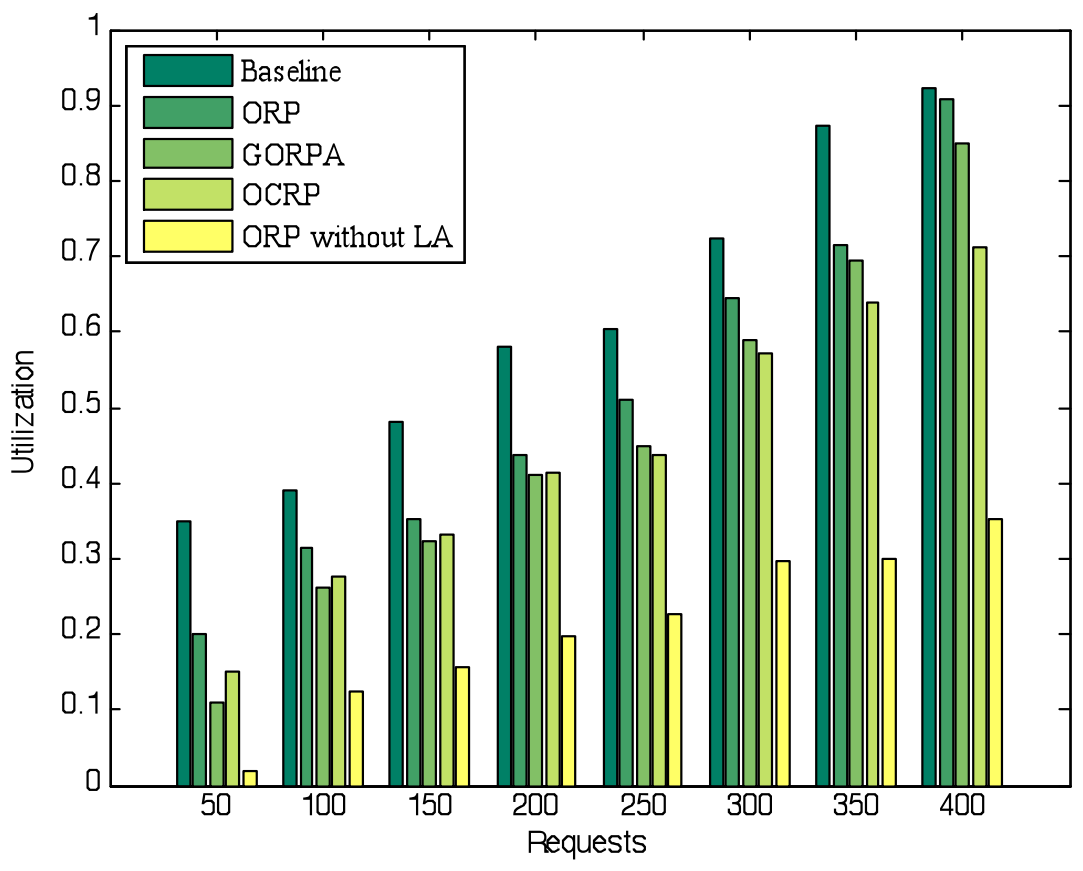

**Fig. 8** Comparison between utilization of VMs

Fig. 8 demonstrates that the utilization of ORP approach is more efficient than the utilization of others; the reason is that ORP is effective in all properties of a VM such as CPU cycle and memory allocation with high resource utilization; Provisioning System in ORP chooses a VM with the most adaption with the service requirements, in each iteration as introduced in Eq. (5). None of other approaches take care of attributes of VMs in addition to the hour costs of VMs while allocation a VM to a service in the request.

The last metric considered in our experiments is the total cost of the selected VMs for deploying the requests. The comparisons of costs of provisioned VMs are performed between ORP, and OCRP [10] and GORPA [4]. Although we used the attributes of VMs presented in Table 2 for the previous experiments, for cost comparison in this experiment, the prices introduced in [10], are applied to compare the techniques in same conditions. The comparison of the cost is generated by Comparison Unit of simulation model depicted in Fig. 4; the results are represented in Fig. 9. It is obvious from the figure that our proposed resource provisioning approach, ORP, obtains better costs comparing with both OCRP and GORPA. The increase of costs with the growth of the requests is expected, which can be seen in Fig. 9. While the requests increase the differences of the total costs of ORP with the others are decreased; the reason is that ORP chooses VMs without considering any future model of requests, and the provisioning is performed in a way that the most proper VMs would be chosen based on the current requests. Therefore the provider must buy new resources and for this the provider may incur additional costs such as costs of new virtual resources offered by IaaS providers and costs of time waiting for the preparation of new VMs [4]. Thus, total costs might have higher prices for new requests as depicted in Fig. 9.

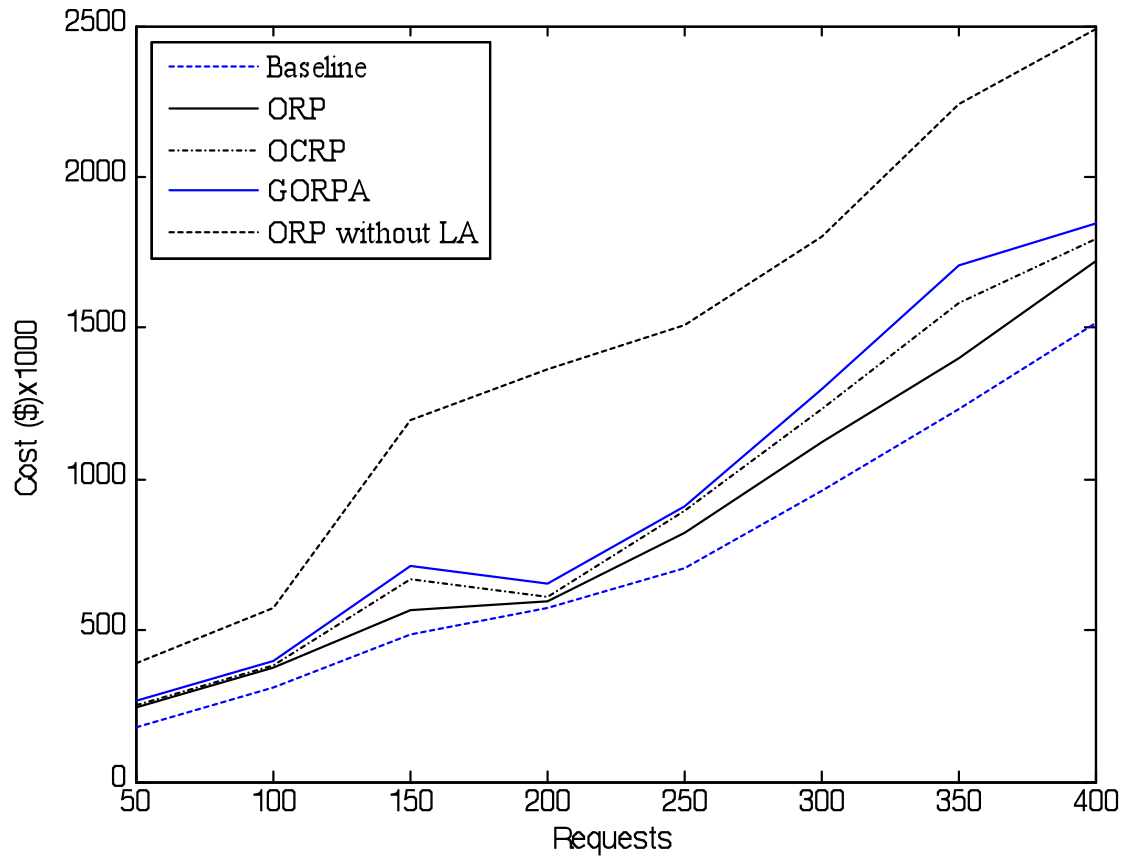

**Fig. 9** Cost comparison

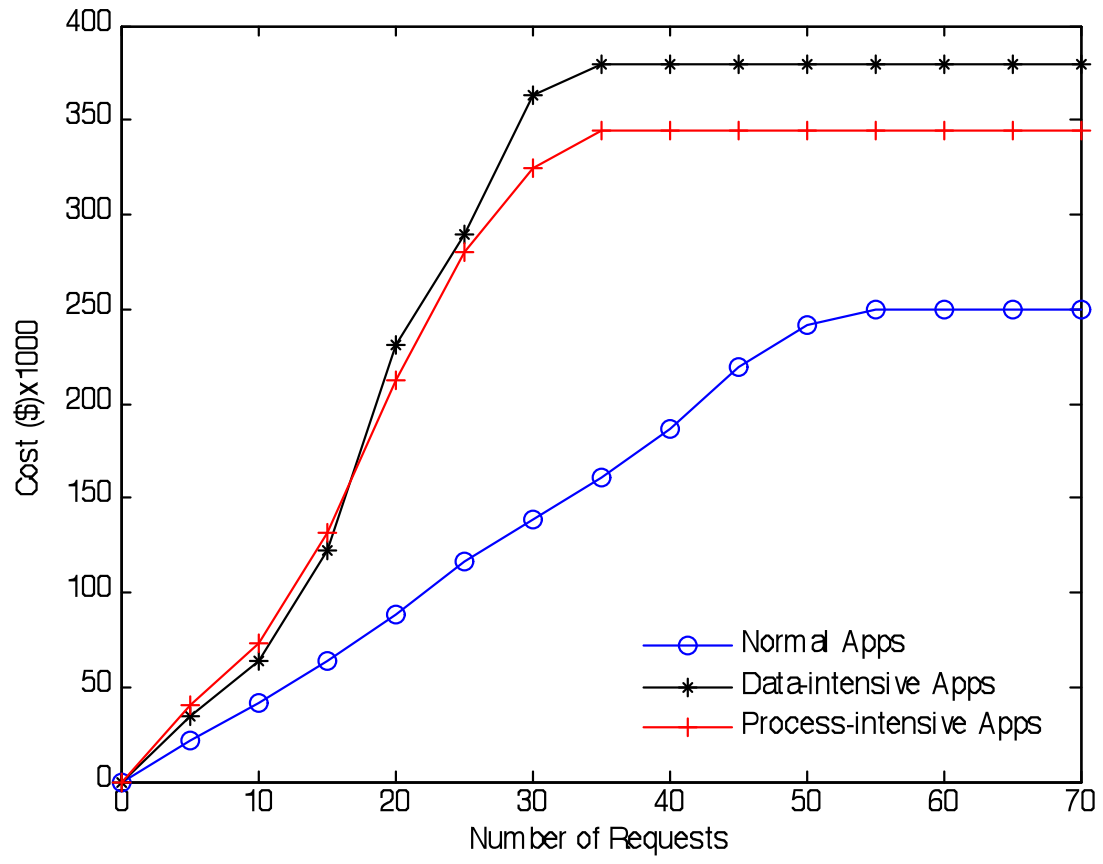

**Fig. 10** Evolution of total costs in different types of demands with constant number of VMs

#### *4.2.2 Impact of ORP on typical application types: data-intensive, process-intensive and normal applications*

In this section, the resource provisioning method has been evaluated for typical application types. We consider three general types of requests, i.e. the requests of data-intensive applications, the requests of process-intensive applications and requests of normal applications. Specifically, we want to show that our model can effectively assign proper VMs to each type of requests privately, with respect to the application requirements, while keeping the costs low. Firstly, the requirements of each application type are discussed; then, a simple scenario to evaluate the performance of ORP based on the type of the requests is presented.

For each type of application, a set of services with different requirements is needed. The applications which devote most of their execution time to computational requirements are deemed process-intensive, whereas applications which require large volumes of data and devote most of their processing time to I/O and manipulation of data are deemed data-intensive. Normal applications have both requirements of data processing and computational processes. Our traced workloads of these types are presented in Table 4.

Table 4
Properties of requests used for experiments of Fig. 7

| Request | Services | VCPU | Memory (GB) | Storage (GB) |
|---|---|---|---|---|
| Data-intensive | $VMSrv_1$ | 1 | 15 | 2×40 |
| | $VMSrv_2$ | 1 | 30 | 1×32 |
| | $VMSrv_3$ | 2 | 60 | 1×80 |
| Process-intensive | $VMSrv_1$ | 4 | 2 | 1×4 |
| | $VMSrv_2$ | 8 | 4 | 1×4 |
| | $VMSrv_3$ | 8 | 8 | 2×16 |
| Normal | $VMSrv_1$ | 1 | 4 | 1×4 |
| | $VMSrv_2$ | 2 | 8 | 1×32 |
| | $VMSrv_3$ | 4 | 15 | 2×80 |

In Fig. 10, the evolutions of total number of requests that are processed by ORP are presented. As expected, in demands of normal applications the growth of costs is smoother while in data-intensive and process-intensive demands, the costs increase faster. Furthermore, since we consider that the virtual resources capacity of each provider is limited, the amount of VMs that each provider dedicates to the requests is bounded; this makes the total cost stop increasing after processing a number of requests. In Fig.10, more requests can be processed while requests are normal application types in comparison with two other types. The reason is that in two later cases, the proper VMs for data-intensive and process-intensive applications finish before in normal application types demands, since special VMs are required in these two types. Therefore, also more increase of costs is seen in data-intensive and process-intensive demands in comparison with normal demands, the number of processed requests decrease as well.

## 5. Conclusion

Cloud computing has enabled new technologies to Software-as-a-Service (SaaS) providers and Infrastructure-as-a-Service (IaaS) providers to offer applications online with pay per use model. These technologies make computing resources more powerful, and thus more efficient resource provisioning techniques must be involved. Current researches of resource provisioning approaches lacks of applications granularity; in this paper, we have proposed an optimized resource provisioning (ORP) approach in order to provide applications, which consist of different services, to users via virtual resources. Providers try to provide the application with an affordable cost while the performance is satisfying. ORP uses LAs on each provider to deploy each request on the best combination of VMs while saving the infrastructural cost. In this article, our proposed approach focuses on how to lower the resource provisioning cost while not severely degrading the performance metrics of services. A comprehensive evaluation is performed for three typical workloads: data-intensive, process-intensive and normal applications. The experimental results show that ORP efficiently adapts the infrastructural requirements, and the resulting performance meets our design goals as well. In general, averages of utilization and cost were improved; in addition the number of requests which can be processed by ORP are optimized.